\documentstyle[11pt]{article}
\begin{document}

\title{\textbf{Thermodynamical description of interacting entropy-corrected new agegraphic dark energy}}

\author{K. Karami$^{1,2}$\thanks{E-mail: KKarami@uok.ac.ir} , A. Sheykhi$^{3,2}$\thanks{E-mail: sheykhi@mail.uk.ac.ir} ,
M. Jamil$^{4}$\thanks{E-mail: mjamil@camp.nust.edu.pk} , F. Felegary$^{1}$, M.M. Soltanzadeh$^{1}$\\\\
$^{1}$\small{Department of Physics, University of Kurdistan,
Pasdaran St., Sanandaj, Iran}\\$^{2}$\small{Research Institute for
Astronomy $\&$ Astrophysics of Maragha (RIAAM), Maragha, Iran}\\
$^{3}$\small{Department of Physics, Shahid Bahonar University, P.O.
Box 76175, Kerman, Iran}\\
$^{4}$\small{Center for Advanced Mathematics and Physics (CAMP),
National University} \\\small{of Sciences and Technology (NUST),
Islamabad, Pakistan}}

\maketitle

\begin{abstract}
To explain the accelerating universe driven by dark energy, a
so-called ``entropy-corrected new agegraphic dark energy'' (ECNADE),
was recently proposed with the help of quantum corrections to the
entropy-area relation in the framework of loop quantum cosmology.
Using this definition, we study its thermodynamical features
including entropy and energy conservation. We discuss the
thermodynamical interpretation of the interaction between ECNADE and
dark matter in a non-flat universe bounded by the apparent horizon.
We obtain a relation between the interaction term of the dark
components and thermal fluctuation.
\end{abstract}

\noindent{\textbf{PACS numbers:}~~~95.36.+x, 04.60.Pp}\\
\noindent{\textbf{Key words:}~~~Dark energy; Loop quantum gravity}

\newpage
\section{Introduction}
Among the overflowing and complementary candidates to explain the
cosmic acceleration, the agegraphic and new agegraphic dark energy
(DE) models condensate in a class of quantum gravity may have
interesting cosmological consequences. These models take into
account the Heisenberg uncertainty relation of quantum mechanics
together with the gravitational effect in general relativity. The
agegraphic DE (ADE) models assume that the observed DE comes from
the spacetime and matter field fluctuations in the universe
\cite{Cai1,Wei2,Wei1}. Since in ADE model the age of the universe is
chosen as the length measure, instead of the horizon distance, the
causality problem in the holographic DE (HDE) is avoided. The ADE
models have been examined and constrained by various astronomical
observations \cite{age,shey1,shey2,karami2}. Although going along a
fundamental theory such as quantum gravity may provide a hopeful way
towards understanding the nature of DE, it is hard to believe that
the physical foundation of ADE is convincing enough. Indeed, it is
fair to say that almost all dynamical DE models are settled at the
phenomenological level, neither HDE model nor ADE model is
exception. Though, under such circumstances, the models of HDE and
ADE, to some extent, still have some advantage comparing to other
dynamical DE models because at least they originate from some
fundamental principles in quantum gravity.

Besides, in the loop quantum gravity, the entropy-area relation can
modify due to the thermal equilibrium fluctuations and quantum
fluctuations \cite{Zhu,Suj,Rovelli}. The corrected entropy takes the
form \cite{Zhang}
\begin{equation}\label{S}
S=\frac{A}{4G}+{\tilde{\alpha}} \ln {\frac{A}{4G}}+{\tilde{\beta}},
\end{equation}
where ${\tilde{\alpha}}$ and ${\tilde{\beta}}$ are two dimensionless
constants of order unity. Motivated by the corrected entropy-area
relation (1)
 the energy density of the ECNADE was proposed as \cite{Karami0}
\begin{equation}
\rho_{\Lambda} = \frac{3n^2{M_P^2}}{\eta^2} +
\frac{\alpha}{{\eta}^4}\ln{({M_P^2}{\eta}^2)} +
\frac{\beta}{\eta^4},\label{ecnade}
\end{equation}
where $\alpha$ and $\beta$ are dimensionless constants of order
unity. Also the conformal time $\eta$ is given by
\begin{equation}
\eta=\int\frac{{\rm d}t}{a}=\int_0^a\frac{{\rm
d}a}{Ha^2}.\label{eta}
\end{equation}
The motivation idea for taking the energy density of ECNADE in the
form (\ref{ecnade}) comes from the fact that both ADE and HDE models
have the same origin. Indeed, it was argued that the ADE models are
the HDE model with different IR length scales \cite{Myung}. In the
special case  $\alpha=\beta=0$, Eq. (\ref{ecnade}) yields the energy
density of NADE in Einstein gravity \cite{Wei2}.

Our main purpose in this paper is to study thermodynamical picture
of the interaction between dark matter (DM) and ECNADE model for a
universe with spacial curvature. Thermodynamical description of the
interaction (coupling) between HDE and DM has been studied in
\cite{wang3,shey2,Jamil1}. The paper is outlined as follows. In the
next section we consider the thermodynamical picture of the
non-interacting ECNADE in a non-flat universe. In section \ref{Int},
we study the thermodynamical description in the case where there is
an interaction term between the dark components. We also present an
expression for the interaction term in terms of a thermal
fluctuation. We conclude our paper in section \ref{Con}.
\section{Thermodynamical description of the non-interacting ECNADE}\label{NonInt}
We start with the homogeneous and isotropic
Friedmann-Robertson-Walker (FRW) universe which is described by the
line element
\begin{equation}
{\rm d}s^2=-{\rm d}t^2+a^2(t)\left(\frac{{\rm d}r^2}{1-kr^2}+r^2{\rm
d}\Omega^2\right),\label{metric}
\end{equation}
where $a(t)$ is the scale factor and $k$  is the curvature parameter
with $k = -1, 0, 1$ corresponding to open, flat, and closed
universes, respectively. The first Friedmann equation governing the
evolution of the universe is
\begin{equation}
{\textsl{H}}^2+\frac{k}{a^2}=\frac{1}{3M_P^2}~
(\rho_{\Lambda}+\rho_{\rm m}),\label{eqfr}
\end{equation}
where $H=\dot{a}/a$ is the Hubble parameter, $\rho_{m}$ and
$\rho_{\Lambda}$ are the energy density of DM and DE, respectively.
The fractional energy densities are defined as
\begin{equation}
\Omega_{\rm m}=\frac{\rho_{\rm m}}{\rho_{\rm cr}}=\frac{\rho_{\rm
m}}{3M_P^2H^2},~~~~~~\Omega_{\rm
\Lambda}=\frac{\rho_{\Lambda}}{\rho_{\rm
cr}}=\frac{\rho_{\Lambda}}{3M_P^2H^2},~~~~~~\Omega_{k}=\frac{k}{a^2H^2}.
\label{eqomega}
\end{equation}
Thus, we can rewrite the first Friedmann equation as
\begin{equation}
\Omega_{\rm m}+\Omega_{\Lambda}=1+\Omega_{k}.\label{eq10}
\end{equation}
From Eq. (\ref{ecnade}) the fractional energy density of the ECNADE
can be written as
\begin{eqnarray}
\Omega_{\Lambda} =
\frac{n^2}{H^2\eta^2}\gamma_{n},\label{density-nade-omega}
\end{eqnarray}
where
\begin{eqnarray}
\label{gamma-parameter1} \gamma_n = 1 +
\frac{1}{3n^2{M_P^2}\eta^2}\Big[\alpha\ln{({M_P^2}{\eta}^2)}
+\beta\Big].
\end{eqnarray}
The continuity equations for DE and pressureless DM are
\begin{equation}
\dot{\rho}_{\Lambda}^0+3H_0(1+\omega_{\Lambda}^{0})\rho_{\Lambda}^0=0,\label{rholamb}
\end{equation}
\begin{equation}
\dot{\rho}_{\rm m}^0+3H_0\rho_{\rm m}^0=0,\label{rhom}
\end{equation}
where $\omega_\Lambda^0$ is the equation of state (EoS) parameter of
ECNADE when it evolves independently of DM. The
superscript/subscript ``0" denotes that there is no interaction
between the dark components and in this picture our universe is in a
thermodynamical stable equilibrium.

The equation of motion of $\Omega_{\Lambda}^0$  is  \cite{Karami0}
\begin{eqnarray}
\label{omegaD-eq-motion2} \Omega^{'0}_{\Lambda} = \Omega_{\Lambda}^0
\Big[3(1 -
\Omega_{\Lambda}^0)+\Omega_k^0+\frac{2}{na_0}\Big({\frac{\Omega_{\Lambda}^0}{\gamma_n^0}}\Big)^{1/2}\Big(2\gamma_n^0
-1-\frac{\alpha
H_0^2}{3{M^{2}_P}n^4}\frac{\Omega_{\Lambda}^0}{\gamma_n^0}\Big)\Big(\frac{\Omega_{\Lambda}^0-1}{\gamma_n^0}\Big)\Big],
\end{eqnarray}
where the prime stands for the derivative with respect to
$x^0=\ln{a_0}$. Taking the time derivative of Eq. (\ref{ecnade}) and
using Eq. (\ref{density-nade-omega}) we get
\begin{equation}\label{dotrho}
\dot{\rho}_{\Lambda}^0=-\frac{2H_0}{na_0\gamma_n^0}\Big({\frac{\Omega_{\Lambda}^0}{\gamma_n^0}}\Big)^{1/2}
\Big(2\gamma_n^0-1-\frac{\alpha
H_0^2}{3{M^{2}_P}n^4}\frac{\Omega_{\Lambda}^0}{\gamma_n^0}\Big)\rho_{\Lambda}^0.
\end{equation}
Substituting into Eq. (\ref{rholamb}) we easily find the EoS
parameter of the non-interacting ECNADE
\begin{equation}
1+\omega_\Lambda^0
=\frac{2}{3na_0\gamma_n^0}\Big({\frac{\Omega_{\Lambda}^0}{\gamma_n^0}}\Big)^{1/2}\Big(2\gamma_n^0
- 1-\frac{\alpha
H_0^2}{3{M^{2}_P}n^4}\frac{\Omega_{\Lambda}^0}{\gamma_n^0}\Big).\label{omegalamb1}
\end{equation}
We also assume the local equilibrium hypothesis to be hold. This
requires that the temperature $T$ of the energy content inside the
apparent horizon should be in equilibrium with the temperature $T_h$
associated with the apparent horizon, so we have $T=T_h$. If the
temperature of the fluid differs much from that of the horizon,
there will be spontaneous heat flow between the horizon and the
fluid and the local equilibrium hypothesis will no longer hold. This
is also at variance with the FRW geometry. Thus, when we consider
the thermal equilibrium state of the universe, the temperature of
the universe is associated with the horizon temperature. The
equilibrium entropy of the ECNADE is connected with its energy and
pressure through the first law of thermodynamics
\begin{equation}
T_0{\rm d}S_\Lambda^0={\rm d}E_\Lambda^0+P_\Lambda^0 {\rm
d}V_0,\label{dS1}
\end{equation}
where the volume enveloped by the apparent horizon is given by
\begin{equation}
V_0=\frac{4\pi}{3}(r_A^0)^3,\label{V}
\end{equation}
and $r_A^0$ is the apparent horizon radius of the FRW universe
\begin{equation}
r_A^0=\frac{1}{\sqrt{H_0^2+k/a_0^2}}.\label{rA}
\end{equation}
The equilibrium energy of the ECNADE inside the apparent horizon is
\begin{equation}
E_\Lambda^0=\rho_\Lambda^0 V_0=\Big[3n^2M_P^2\eta_0^{-2}+\alpha
\eta_0^{-4}\ln(M_P^2\eta_0^{2})+\beta
\eta_0^{-4}\Big]\Big[\frac{4\pi}{3}(r_A^0)^3\Big].\label{E}
\end{equation}
Taking the differential form of Eq. (\ref{E}) and using Eq.
(\ref{density-nade-omega}), we find
\begin{equation}
{\rm d}E_\Lambda^0=-12\pi
(r_A^0)^3M_P^2H_0^2\Omega_\Lambda^0\Big[(1+\omega_\Lambda^0){\rm
d}x^0-\frac{{\rm d}r_A^0}{r_A^0}\Big].\label{dE}
\end{equation}
The associated temperature on the apparent horizon can be written as
\begin{equation}
T_0=\frac{1}{2\pi r_A^0}.\label{T0}
\end{equation}
Finally, combining Eqs. (\ref{dE}) and (\ref{T0}) with Eq.
(\ref{dS1}) we obtain
\begin{equation}
{\rm
d}S_\Lambda^0=24\pi^2(r_A^{0})^3M_P^2H_0^{2}\Omega_\Lambda^0(1+\omega_\Lambda^0)\Big[{\rm
d}r_A^0-r_A^0{\rm d}x^0\Big].\label{dSlamb0}
\end{equation}
\section{Thermodynamical description of the interacting ECNADE}\label{Int}
Next we generalize our study to the case where the pressureless DM
and the ECNADE interact with each other. In this case $\rho_{m}$ and
$\rho_{\Lambda}$ do not conserve separately; they must rather enter
the energy balances \cite{jamil}
\begin{equation}
\dot{\rho}_{\Lambda}+3H(1+\omega_{\Lambda})\rho_{\Lambda}=-Q,\label{eqpol}
\end{equation}
\begin{equation}
\dot{\rho}_{\rm m}+3H\rho_{\rm m}=Q,\label{eqCDM}
\end{equation}
where $Q$  is the interaction term. Inserting Eq. (\ref{dotrho})
without superscript/subscript ``0" into (\ref{eqpol}), we obtain the
EoS parameter of the interacting ECNADE
\begin{equation}
1+\omega_{\Lambda}=\frac{2}{3na\gamma_n}\Big({\frac{\Omega_{\Lambda}}{\gamma_n}}\Big)^{1/2}\Big(2\gamma_n
- 1-\frac{\alpha
H^2}{3{M^{2}_P}n^4}\frac{\Omega_{\Lambda}}{\gamma_n}\Big)-\frac{Q}{9M_p^2H^3\Omega_\Lambda}.\label{omegalamb1}
\end{equation}
The evolution behavior of the ECNADE is now given by \cite{Karami0}
\begin{eqnarray}
\label{omegaD-eq-motion2} {\Omega^{'}_{\Lambda}} = \Omega_{\Lambda}
\Big[3(1 - \Omega_{\Lambda})  - 3b^2(1 + \Omega_k) +
\Omega_k~~~~~~~~~~~~~~~~~~~~~~~~~~~~~~~~~~~~~~~~~\nonumber\\+\frac{2}{na}\Big({\frac{\Omega_{\Lambda}}{\gamma_n}}\Big)^{1/2}\Big(2\gamma_n
-1-\frac{\alpha
H^2}{3{M^{2}_P}n^4}\frac{\Omega_{\Lambda}}{\gamma_n}\Big)\Big(\frac{\Omega_{\Lambda}-1}{\gamma_n}\Big)\Big].~~~
\end{eqnarray}
As soon as an interaction between dark components is taken into
account, they cannot remain in their respective equilibrium states.
The effect of interaction between the dark components is
thermodynamically interpreted as a small fluctuation around the
thermal equilibrium. Therefore, the entropy of the ECNADE is
connected with its energy and pressure through the first law of
thermodynamics
\begin{equation}
T{\rm d}S_\Lambda={\rm d}E_\Lambda+P_\Lambda {\rm d}V,\label{S}
\end{equation}
where
\begin{equation}
T=\frac{1}{2\pi r_A},
\end{equation}
\begin{equation}
V=\frac{4\pi}{3}r_A^3,
\end{equation}
and
\begin{equation}
r_A=\frac{1}{\sqrt{H^2+k/a^2}}.
\end{equation}
Now we have an extra logarithmic correction term in the entropy
expression
\begin{equation}
S_\Lambda=S_\Lambda^{(0)}+S_\Lambda^{(1)},\label{Slamb}
\end{equation}
where
\begin{equation}
S_\Lambda^{(1)}=-\frac{1}{2}\ln(CT_0^2),\label{S1}
\end{equation}
is the leading logarithmic correction and $C$ is the heat capacity
defined as
\begin{equation}
C=T_0\frac{\partial S_\Lambda^{(0)}}{\partial T_0}.\label{C}
\end{equation}
It is a matter of calculation to show that
\begin{equation}
C=-24\pi^2(r_A^{0})^4M_P^2H_0^2\Omega_\Lambda^{0}(1+\omega_\Lambda^{0}).\label{C1}
\end{equation}
In addition we get
\begin{equation}
S_\Lambda^{(1)}=-\frac{1}{2}\ln
\Big[-6(r_A^0)^2M_P^2H_0^2\Omega_\Lambda^0(1+\omega_\Lambda^0)\Big].\label{Slamb1}
\end{equation}
It is easy to show that
\begin{equation}
{\rm d}S_\Lambda=24\pi^2r_A^3M_P^2H^2\Omega_\Lambda
\Big[(1+\omega_\Lambda){\rm
d}r_A-\frac{2r_A}{3na\gamma_n}\Big({\frac{\Omega_{\Lambda}}{\gamma_n}}\Big)^{1/2}\Big(2\gamma_n
- 1-\frac{\alpha
H^2}{3{M^{2}_P}n^4}\frac{\Omega_{\Lambda}}{\gamma_n}\Big){\rm
d}x\Big],\label{dS1-v1}
\end{equation}
where
\begin{eqnarray}
1+\omega_\Lambda=\frac{1}{24\pi^2r_A^3M_P^2H^2\Omega_\Lambda}\Big(\frac{{\rm
d}S_\Lambda^{(0)}}{{\rm d}r_A}+\frac{{\rm d}S_\Lambda^{(1)}}{{\rm
d}r_A}\Big)~~~~~~~~~~~~~~~\nonumber\\+\frac{2r_A}{3na\gamma_n}\Big({\frac{\Omega_{\Lambda}}{\gamma_n}}\Big)^{1/2}\Big(2\gamma_n
- 1-\frac{\alpha
H^2}{3{M^{2}_P}n^4}\frac{\Omega_{\Lambda}}{\gamma_n}\Big)\frac{{\rm
d}x}{{\rm d}r_A}.\label{omegalamb2-v1}
\end{eqnarray}
Substituting the expressions for the volume, energy, and temperature
in Eq. (\ref{S}) for the interacting case, we get
\begin{equation}
{\rm d}S_\Lambda=-24\pi^2r_A^4M_P^2H^2\Omega_\Lambda
\Big[(1+\omega_\Lambda)\Big({\rm d}x-\frac{{\rm
d}r_A}{r_A}\Big)+\frac{Q}{9M_P^2H^3\Omega_\Lambda}{\rm
d}x\Big],\label{dS1-v2}
\end{equation}
where
\begin{equation}
1+\omega_\Lambda=-\Big[\frac{1}{24\pi^2r_A^4M_P^2H^2\Omega_\Lambda}\Big(\frac{{\rm
d}S_\Lambda^{(0)}}{{\rm d}r_A}+\frac{{\rm d}S_\Lambda^{(1)}}{{\rm
d}r_A}\Big)+\frac{Q}{9M_P^2H^3\Omega_\Lambda}\frac{{\rm d}x}{{\rm
d}r_A}\Big]\Big(\frac{{\rm d}x}{{\rm
d}r_A}-\frac{1}{r_A}\Big)^{-1}.\label{omegalamb2-v2}
\end{equation}
Using Eq. (\ref{dSlamb0}) one finds
\begin{equation}
\frac{{\rm d}S_\Lambda^{(0)}}{{\rm d}r_A}=\frac{\partial
S_\Lambda^{(0)}}{\partial r_A^0}\frac{{\rm d}r_A^0}{{\rm
d}r_A}+\frac{\partial S_\Lambda^{(0)}}{\partial x^0}\frac{{\rm
d}x^0}{{\rm d}r_A},\label{dS0}
\end{equation}
where
\begin{equation}
\frac{{\rm d}r_A^0}{{\rm
d}x^0}=\frac{1+\Omega_k^0+q_0}{H_0(1+\Omega_k^0)^{3/2}},
\end{equation}
\begin{equation}
\frac{{\rm d}r_A^0}{{\rm d}r_A}=\frac{{\rm d}r_A^0/{\rm d} t}{{\rm
d}r_A/{\rm d} t}=\frac{H_0{\rm d}r_A^0/{\rm d} x^0}{H{\rm d}r_A/{\rm
d} x}=\Big(\frac{1+\Omega_k^0+q_0}{1+\Omega_k+q}\Big)\Big(
\frac{1+\Omega_k}{1+\Omega_k^0}\Big)^{3/2},
\end{equation}
\begin{equation}
\frac{{\rm d}x^0}{{\rm d}r_A}=\frac{{\rm d}x^0/{\rm d} t}{{\rm
d}r_A/{\rm d} t}=\frac{H_0{\rm d}x^0/{\rm d} x^0}{H{\rm d}r_A/{\rm
d} x}=\frac{H_0(1+\Omega_k)^{3/2}}{1+\Omega_k+q},
\end{equation}
and $q=-1-H^{-1}{\rm d}H/{\rm d}x$ is the deceleration parameter.
Thus we can rewrite Eq. (\ref{dS0}) as
\begin{equation}
\frac{{\rm d}S_\Lambda^{(0)}}{{\rm
d}r_A}=24\pi^2(r_A^0)^3M_P^2H_0^2\Omega_\Lambda^0\left[\frac{(1+\omega_\Lambda^0)(1+\Omega_k)^{3/2}}{1+\Omega_k+q}\right]
\left[\frac{1+\Omega_k^0+q_0}{(1+\Omega_k^0)^{{3/2}}}-H_0r_A^0\right].\label{ds00}
\end{equation}
In a similar manner we have
\begin{equation}
\frac{{\rm d}S_\Lambda^{(1)}}{{\rm
d}r_A}=-\frac{1}{2}\frac{(1+\Omega_k)^{{3/2}}}{(1+\Omega_k+q)}\frac{{\rm
d}}{{\rm
d}t}\ln\Big[(r_A^0)^2H_0^2\Omega_\Lambda^{0}(1+\omega_\Lambda^{0})\Big].\label{ds111}
\end{equation}
Therefore, the interaction term may be obtained as
\begin{equation}
\frac{Q}{9M_P^2H^3\Omega_\Lambda}=(1+\omega_\Lambda)\left(\frac{1}{r_A}\frac{{\rm
d}r_A}{{\rm d}x}-1\right)-\left(\frac{\frac{{\rm d}r_A}{{\rm
d}x}}{24\pi^2r_A^4M_P^2H^2\Omega_\Lambda}\right)\left(\frac{{\rm
d}S_\Lambda^{(0)}}{{\rm d}r_A}+\frac{{\rm d}S_\Lambda^{(1)}}{{\rm
d}r_A}\right),\label{Q}
\end{equation}
 which can also be written as
\begin{eqnarray}
\frac{Q}{9M_P^2H^3\Omega_\Lambda}=\frac{q(1+\omega_\Lambda)}{1+\Omega_k}
-\frac{(r_A^{0})^3H_0^2\Omega_\Lambda^{0}(1+\omega_\Lambda^{0})q_0}{r_A^4H^3\Omega_\Lambda(1+\Omega_k^0)^{{3/2}}}~~~~~~~~~~~~~~~
\nonumber\\+\frac{1}{48\pi^2r_A^4M_P^2H^3\Omega_\Lambda}\frac{{\rm
d}}{{\rm
d}t}\ln\Big[(r_A^0)^2H_0^2\Omega_\Lambda^{0}(1+\omega_\Lambda^{0})\Big].\label{Q1}
\end{eqnarray}
In this way we provide the relation between the interaction term of
the dark components and the thermal fluctuation.
\section{Concluding remarks}\label{Con}

In this paper, we studied the model of interacting ECNADE by
considering the entropy corrections to the NADE model. These
corrections are motivated from the loop quantum cosmology which is
one of the promising theories of quantum gravity. We restricted our
study to the leading order correction which contains the logarithmic
of the area, however, one can improve this study by considering
higher order corrections to get better thermodynamic interpretation.
We provided a thermodynamical description of the ECNADE model in a
universe with spacial curvature. We assumed that in the absence of a
coupling, the two dark components remain in separate thermal
equilibrium with the horizon and that the presence of a small
coupling between them can be described as stable fluctuations around
equilibrium. Finally, resorting to the logarithmic correction to the
equilibrium entropy we derived an expression for the interaction
term in terms of a thermal fluctuation.
\\
\\
\noindent{\textbf{Acknowledgements}\\
The works of K. Karami and A. Sheykhi have been supported
financially by Research Institute for Astronomy $\&$ Astrophysics of
Maragha (RIAAM), Maragha, Iran.


\end{document}